\documentclass[aps,prl,twocolumn,showpacs,superscriptaddress,groupedaddress]{revtex4}  
\usepackage{graphicx}  
\usepackage{dcolumn}   
\usepackage{bm}        
\usepackage{amssymb}   
\usepackage{color}
\usepackage{amsmath}
\begin{document}

\widetext

\begin{flushright}
IPM/P-2018/026\\
\end{flushright}

\title{
Entanglement Renormalization for Weakly Interacting Fields}

\author{Jordan S. Cotler}\email{jcotler@stanford.edu}

\affiliation{\it Stanford Institute for Theoretical Physics, Stanford University, Stanford, CA 94305, US}

\author{M. Reza Mohammadi Mozaffar} \email{m_mohammadi@ipm.ir}

\affiliation{School of Physics, Institute for Research in Fundamental Sciences (IPM), P.O.Box 19395-5531, Tehran, Iran}

\author{Ali Mollabashi} \email{mollabashi@ipm.ir}

\affiliation{School of Physics, Institute for Research in Fundamental Sciences (IPM), P.O.Box 19395-5531, Tehran, Iran}

\author{Ali Naseh} \email{naseh@ipm.ir}

\affiliation{School of Particles and Accelerators, Institute for Research in Fundamental Sciences (IPM), P.O.Box 19395-5531, Tehran, Iran}

\date{\today}

\begin{abstract}
We adapt the techniques of entanglement renormalization tensor networks to weakly interacting quantum field theories in the continuum.  A key tool is ``quantum circuit perturbation theory,'' which enables us to systematically construct unitaries that map between wavefunctionals which are Gaussian with arbitrary perturbative corrections. As an application, we construct a local, continuous MERA (cMERA) circuit that maps an unentangled scale-invariant state to the ground state of $\varphi^4$ theory to 1-loop.  Our local cMERA circuit corresponds exactly to 1-loop Wilsonian RG on the spatial momentum modes.  In other words, we establish that perturbative Wilsonian RG on spatial momentum modes can be equivalently recast as a local cMERA circuit in $\varphi^4$ theory, and argue that this correspondence holds more generally.  Our analysis also suggests useful numerical ansatzes for cMERA in the non-perturbative regime.
\end{abstract}

\maketitle
\section{Introduction}

Tensor networks have become a transformative tool for numerically analyzing 1D quantum systems on the lattice, as well as exploring properties of 1D many-body states with area-law entanglement \cite{DMRG1, DMRG2, DMRG3, ER1, ER2, ER3, MPS1, MPS2, MPS3}.  However, there are various obstructions to generalizing tensor networks (i) to higher dimensions and (ii) to continuum field theories.  Tensor networks for higher-dimensional lattices are computationally difficult to implement since accurate numerics can require an intractably large number of tensor contractions, even for systems of modest size \cite{Contract1}.  On the other hand, the predominant continuum tensor network cMPS is useful for non-relativistic systems in 1+1 dimensions, but fails for relativistic systems in 1+1 dimensions and also suffers from the ``contraction problem'' in higher dimensions \cite{Verstraete:2010ft, cMPSrel1, cMPSrel2, cMPSfermionic1}.  

A promising tensor network architecture designed to work both in higher dimensions and in the continuum is called cMERA, the continuum analog of MERA (multi-scale entanglement renormalization ansatz) \cite{cMERA1}.  cMERA generates variational ansatzes which have a hierarchical pattern of entanglement across distance scales.  The construction of the cMERA state is inspired by spatial renormalization group (RG) methods.  Each layer of the network comprising the state corresponds to a step of renormalization group flow.

For all its promise, cMERA has only been constructed for the ground states of free field theories which are solvable using standard methods \cite{cMERA1, cMERA2, cMERA3, cMERA4}.  Even in the context of mean field theory, cMERA has limited utility over known methods \cite{CMM1}.

In this paper, we take the first steps towards applying cMERA to interacting field theories.  In particular, we use quantum circuit perturbation theory to construct a cMERA for the ground state of $\varphi^4$ theory to 1-loop in perturbation theory.  Remarkably, we can construct a \textit{local} cMERA circuit which corresponds exactly to 1-loop Wilsonian RG on spatial momentum modes.  This establishes a connection between cMERA and more conventional forms of RG.  Our perturbative analysis of cMERA leads us to formulate a numerical cMERA ansatz for ground states of interacting theories, which need not have weak coupling.  A companion paper \cite{} contains many techniques, and details of our calculations.

\section{Circuit Perturbation Theory}

Before constructing cMERA circuits for weakly interacting field theories, we need to gain facility with manipulating quantum circuits for QFT's.  Our core tool is ``quantum circuit perturbation theory'' which we summarize here, and develop in detail in \cite{RGcircuits1}.  A central, question is: given two states of a quantum field theory $|\Psi_1\rangle$ and $|\Psi_2\rangle$ which are each Gaussian with arbitrary perturbative corrections, how do we systematically construct a unitary $U$ such that $U|\Psi_1\rangle = |\Psi_2\rangle$?

For concreteness, we restrict our attention to scalar field theory in $d$--dimensions.  The canonical commutation relations are $[\widehat{\phi}(\vec{x}),\widehat{\pi}(\vec{y})] = i \, \delta^d(\vec{x}-\vec{y})$, where we have set $\hbar = 1$.  A Gaussian state $|\Psi\rangle$ has a wavefunctional of the form
\begin{equation}
\label{GaussianState1}
\langle \phi| \Psi\rangle = \mathcal{N} \, e^{-\frac{1}{2}\int d^d \vec{x}\,d^d \vec{y}\, \left(\phi(\vec{x}) - a(\vec{x})\right) \,b(\vec{x},\vec{y})\, \left(\phi(\vec{y}) - a(\vec{y})\right) }
\end{equation}
where $\mathcal{N}$ is an overall normalization.  Throughout the paper, we will use $\mathcal{N}$ as a placeholder for normalization.  We see that the state is completely determined by its one and two-point correlation functions.  We will be primarily interested in \textit{translation} and \textit{rotation-invariant} Gaussian states -- this corresponds to states of the form in Eqn.~\eqref{GaussianState1} for which $a(\vec{x}) = \text{const.}$ and $b(\vec{x},\vec{y}) = b(|\vec{x}-\vec{y}|)$.

In scalar field theory, we can write any non-singular Hermitian operator $O$ as
\begin{equation}
\label{generatorExpand2}
O = \sum_{n=0}^\infty \sum_{k=0}^n \int d^d \vec{x}_1 \cdots d^d \vec{x}_n \, c_{n}^{(k)}(\vec{x}_1,...,\vec{x}_n) \, S_{n}^{(k)}(\vec{x}_1,...,\vec{x}_n)
\end{equation}
where all the $c_n^{(k)}$ are real-valued functions or distributions, and
\begin{align}
\label{fieldbasis1}
S_{n}^{(k)}(\vec{x}_1,...,\vec{x}_n) =&\, \widehat{\phi}(\vec{x}_1)\cdots \widehat{\phi}(\vec{x}_k) \widehat{\pi}(\vec{x}_{k+1}) \cdots \widehat{\pi}(\vec{x}_n) \nonumber \\
& + \widehat{\pi}(\vec{x}_{k+1}) \cdots \widehat{\pi}(\vec{x}_n) \widehat{\phi}(\vec{x}_1)\cdots \widehat{\phi}(\vec{x}_k)
\end{align}
where $1 \leq k \leq n$.  In other words, $\{S_n^{(k)}\}$ generates all operators in the theory.  The quadratic operators, which are generated by $\{S_n^{(k)}\}_{n \leq 2}$\,, have particularly nice properties: they form a (closed) Lie algebra, and generate unitaries which map Gaussian states to Gaussian states.  In terms of equations, if $Q, Q'$ are quadratic operators, then $[Q, Q']$ is also a quadratic operator.  If $|\Psi\rangle$ is a Gaussian state, then $e^{-i Q} |\Psi\rangle$ is also a Gaussian state. 

Given two (translation and rotation-invariant) \textit{Gaussian} states $|\Psi_1^G\rangle, |\Psi_2^G\rangle$, one can systematically find quadratic operators $Q$ such that $e^{-i Q}|\Psi_1^G\rangle = |\Psi_2^G\rangle$ (see \cite{RGcircuits1} for explicit constructions).  This systematic construction is possible due to the technology of squeezed coherent states, which leverages that quadratic operators form a manageably small (closed) Lie algebra.  In contrast, given two \textit{non}-Gaussian states, it is generally \textit{not} possible to systematically construct unitaries which map between the states.  This problem amounts to considering the equation $e^{-i O}|\Psi_1\rangle = |\Psi_2\rangle$ for a generic $O$ (as per Eqn.~\eqref{generatorExpand2}) with undetermined $c_n^{(k)}$'s, and then finding $c_n^{(k)}$'s which satisfy the equation.

Luckily, there is a tractable middle ground between the Gaussian and non-Gaussian cases.  Suppose we have some small parameter $\epsilon$, and that $|\Psi_1\rangle,|\Psi_2\rangle$ are Gaussian up to perturbative corrections in $\epsilon$.  Specifically, suppose that we consider first order corrections in $\epsilon$ of the form
\begin{align}
|\Psi_1\rangle = (1 - i \epsilon R_1)|\Psi_1^G\rangle\,, \quad |\Psi_2\rangle = (1 - i \epsilon R_2)|\Psi_2^G\rangle 
\end{align}
where $R_1, R_2$ are generated by $\{S_n^{(k)}\}_{n \leq N}$ for some $N$, meaning that $R_1, R_2$ do not contain products of $\widehat{\phi}$'s and $\widehat{\pi}$'s that exceed length $N$.  The analysis that follows generalizes to arbitrary orders in $\epsilon$.

We will construct a unitary of the form $U = e^{-i(Q + \epsilon R)}$ where $R$ is generated by $\{S_n^{(k)}\}_{n \leq N}$, such that $U|\Psi_1\rangle = |\Psi_2\rangle + \mathcal{O}(\epsilon^2)$.  Let $Q$ be a quadratic operator which satisfies $e^{-i Q}|\Psi_1^G\rangle = |\Psi_2^G\rangle$.  Given $|\Psi_1^G\rangle,|\Psi_2^G\rangle$, we can construct such a $Q$ explicitly.  Then using various manipulations of the Baker-Campbell-Hausdorff formula, we obtain
\begin{align}
\label{Req1}
R &= \frac{i\,\text{ad}_{Q}}{1 - e^{i \,\text{ad}_{Q}}}\, R_1 + \frac{i\,\text{ad}_{Q}}{1 - e^{- i \,\text{ad}_{Q}}} \, R_2 
\end{align}
where the superoperator $\text{ad}_A$ acts by $\text{ad}_A B = [A,B]$.  Even though Eqn.~\eqref{Req1} may appear unwieldy -- when we expand out the power series in the $\text{ad}_Q$ operators, we find an infinite sum of nested commutators -- there is a crucial simplification: The commutator of a quadratic operator with \textit{any} operator generated by $\{S_n^{(k)}\}_{n \leq N}$ yields another operator which is still in $\{S_n^{(k)}\}_{n \leq N}$.  As a consequence, we can write $R$ above as $R = R_1' + R_2'$ where $R_1', R_2'$ are each in $\{S_n^{(k)}\}_{n \leq N}$.  Furthermore, given specific states $|\Psi_1\rangle, |\Psi_2\rangle$ which are translation and rotation-invariant, we can \textit{explicitly compute} $Q$ \textit{as well as} $R$, namely by explicitly evaluating Eqn.~\eqref{Req1}. 

Said in a different way, manipulations of quadratic operators are tractable because $\{S_n^{(k)}\}_{n \leq 2}$ forms a basis for a (closed) Lie algebra.  By contrast, non-quadratic operators are harder to handle because $\{S_n^{(k)}\}_{n \leq N}$ for any $N > 2$ does \textit{not} form a basis for a closed Lie algebra, since commutators of non-quadratic operators generically yield operators with progressively longer products of $\widehat{\phi}$'s and $\widehat{\pi}$'s.  However, $\{S_n^{(k)}\}_{n \leq 2} \cup \{\epsilon \, S_n^{(k)}\}_{n \leq N}$ \textit{does} form a basis for a closed Lie algebra to $\mathcal{O}(\epsilon)$, which enables us to evaluate various sums of nested commutators (such as in Eqn.~\eqref{Req1}) in closed form.  At higher orders in $\epsilon$, we would leverage the fact that
\begin{equation}
\{S_n^{(k)}\}_{n \leq 2} \cup \{\epsilon \, S_n^{(k)}\}_{n \leq N} \cup \bigcup_{\ell=2}^m \{\epsilon^\ell \, S_n^{(k)}\}_{n \leq \ell(N-1)}
\end{equation}
forms a basis for a closed Lie algebra to $\mathcal{O}(\epsilon^m)$.

In the next section, we construct a circuit from an arbitrary scale-invariant, zero-mean, Gaussian wavefunctional to the ground state of $\varphi^4$ theory at 1-loop in the perturbative coupling.  In subsequent sections, we will use this result to construct a \textit{local} position-space cMERA which is equivalent to 1-loop Wilsonian RG on spatial momentum modes.

\section{1-loop circuit from Gaussian to Ground State of $\varphi^4$ theory}

It will be convenient for us to work in momentum space.  We choose Fourier conventions so that $[\widehat{\phi}(\vec{k}),\widehat{\pi}(\vec{p})] = i \, \delta^d(\vec{k}+\vec{p})$.  The Hamiltonian for $\varphi^4$ theory is given by
\begin{align}
& H^\Lambda = \frac{1}{2} \int^\Lambda d^d \vec{k} \, \left(\widehat{\pi}_{\vec{k}} \, \widehat{\pi}_{-\vec{k}} + \widehat{\phi}_{\vec{k}} \left(\vec{k}^2 + m^2\right) \widehat{\phi}_{-\vec{k}}\right) \nonumber \\
& \qquad + \frac{\lambda}{4!}\frac{1}{(2\pi)^{d}}\int^\Lambda d^d \vec{k}_1 \, d^d \vec{k}_2 \, d^d \vec{k}_3 \, \widehat{\phi}_{\vec{k}_1} \widehat{\phi}_{\vec{k}_2} \widehat{\phi}_{\vec{k}_3} \widehat{\phi}_{-\vec{k}_1 - \vec{k}_2 - \vec{k}_3}
\end{align}
where $\widehat{\phi}_{k} := \widehat{\phi}(\vec{k})$, $\widehat{\pi}_{k} := \widehat{\pi}(\vec{k})$, and we have imposed a UV cutoff at momentum scale $|\vec{k}| = \Lambda$.  Next we will renormalize the Hamiltonian to scale $\Lambda e^u$, where $-\infty < u \leq 0$.  After performing $1$-loop Wilsonian renormalization on the spatial momentum modes, we obtain
\begin{align}
\label{phi4Ham1loop}
&H_{1-\text{loop}}^{\Lambda e^u} = \frac{1}{2} \int^\Lambda d^d \vec{k} \, \left(\widehat{\pi}_{\vec{k}}\,\widehat{\pi}_{-\vec{k}} + \widehat{\phi}_{\vec{k}} \left(\vec{k}^2 + e^{-2u}\,\widetilde{m}^2\right) \widehat{\phi}_{-\vec{k}}\right) \nonumber \\
& \,\,+ \frac{e^{(d-3)u}\lambda}{4!}\frac{1}{(2\pi)^{d}} \int^\Lambda d^d \vec{k}_1 \, d^d \vec{k}_2 \, d^d \vec{k}_3 \, \widehat{\phi}_{\vec{k}_1} \widehat{\phi}_{\vec{k}_2} \widehat{\phi}_{\vec{k}_3} \widehat{\phi}_{-\vec{k}_1 - \vec{k}_2 - \vec{k}_3}
\end{align}
where $\widetilde{m}^2$ is given by
\begin{align}
\label{deltaM1}
\widetilde{m}^2 &= m^2 + \frac{\lambda}{2} \int_{\Lambda e^u}^\Lambda \frac{d^{d}\vec{k}}{(2\pi)^d}  \frac{1}{\vec{k}^2+ m^2}=: m^2 + \delta m^2\,.
\end{align}
Using $H_{1-\text{loop}}^{\Lambda e^u}$\,, we calculate the ground state wavefunctional of $\varphi^4$ theory at scale $\Lambda e^u$ to $1$-loop \cite{Hatfield1}:
\begin{align}
\label{phi4gs1}
\langle \phi | \Psi(\Lambda e^u)\rangle &= \mathcal{N} \, e^{- G[\phi] - e^{-2u} \delta m^2 \, R_1[\phi] - e^{(d-3)u} \lambda \, R_2[\phi]} + \mathcal{O}(\lambda^2)
\end{align}
where $G[\phi]$, $R_1[\phi]$, $R_2[\phi]$ are given by
\begin{align}
G[\phi] &= \frac{1}{2}\int^\Lambda d^d \vec{k} \, \phi_{\vec{k}} \,\,  \omega_k \, \phi_{-\vec{k}}
\\
R_1[\phi]&=\frac{1}{4}\int^\Lambda d^d \vec{k}\, \frac{1}{\omega_k}\,\phi_{\vec{k}} \,\phi_{-\vec{k}}
\end{align}
\begin{align}
\begin{split}
R_2[\phi] &=\frac{1}{16}\int^\Lambda d^d \vec{k}\, \frac{1}{\omega_k}\,\left(\int \frac{d^d \vec{q}}{(2\pi)^{d}} \frac{1}{\omega_k + \omega_q} \right)\,\phi_{\vec{k}}\, \phi_{-\vec{k}}
\\
& \quad + \frac{1}{24}\frac{1}{(2\pi)^d} \int^\Lambda \frac{d^d \vec{k}_1 \, d^d \vec{k}_2 \, d^d \vec{k}_3}{\omega_{k_1}\!+\!\omega_{k_2}\!+\!\omega_{k_3}\!+\!\omega_{-\vec{k}_1 - \vec{k}_2 - \vec{k}_3}} \\
& \qquad \qquad \qquad \qquad \qquad \times \phi_{\vec{k}_1} \phi_{\vec{k}_2} \phi_{\vec{k}_3} \phi_{-\vec{k}_1 - \vec{k}_2 - \vec{k}_3}\,,
\end{split}
\end{align}
where $\omega_k := \sqrt{\vec{k}^2 + e^{-2u} m^2}$.
Next we introduce a reference Gaussian state $|\Psi_0\rangle$, namely
\begin{equation}
\label{wavefunctional0momen1}
\langle \phi |\Psi_0\rangle = \text{det}^{\frac{1}{4}}\left(\frac{\Omega}{\pi}\right) \, \exp\left(-\frac{1}{2} \int d^d \vec{k} \, \phi_{\vec{k}} \, \Omega(\vec{k}) \, \phi_{-\vec{k}} \right)\,,
\end{equation}
\vskip.1cm
\noindent which is translation and rotation-invariant (i.e., $\Omega(\vec{k}) = \Omega(|\vec{k}|)$\,) and has zero mean.  The kernel in the exponent of Eqn.~\eqref{wavefunctional0momen1} is related
to the inverse equal-time Green's function, since $\langle \Psi_0| \,\widehat{\phi}(\vec{p}) \, \widehat{\phi}(\vec{k}) \, |\Psi_0\rangle = \frac{1}{2 \Omega(\vec{k})} \, \delta^{(d)}(\vec{p} + \vec{k})$.

We construct a unitary $U$ such that
\begin{equation}
U|\Psi_0\rangle = |\Psi(\Lambda e^u)\rangle \,\,+ \mathcal{O}(\lambda^2)\,.
\end{equation}
Our unitary is
\begin{align}
\label{unitary1}
U &= \exp\left(i \, K_{2,0} + i \lambda \, (K_{2,1} + K_4)\right)
\end{align}
with
\begin{widetext}
\vskip-.7cm
\begin{align}
K_{2,0} &= - \int d^d \vec{k}_1 \, d^d \vec{k}_2 \, \delta^{(d)}(\vec{k}_1 + \vec{k}_2)\,g_{2,0}(\vec{k}_1) \, S_2^{(1)}(\vec{k}_1, \vec{k}_2), \\
K_{2,1} &= - \int d^d \vec{k}_1 \, d^d \vec{k}_2 \,\delta^{(d)}(\vec{k}_1 + \vec{k}_2)\, g_{2,1}(\vec{k}_1) \, S_2^{(1)}(\vec{k}_1, \vec{k}_2), \\
K_{4} &= \int d^d \vec{k}_1 \, d^d \vec{k}_2 \, d^d \vec{k}_3 \, d^d \vec{k}_4 \, \delta^{(d)}(\vec{k}_1 + \vec{k}_2 + \vec{k}_3 + \vec{k}_4) \bigg( g_{4}^{(1)}(\vec{k}_1,\vec{k}_2,\vec{k}_3,\vec{k}_4) \, S_4^{(1)}(\vec{k}_1, \vec{k}_2, \vec{k}_3, \vec{k}_4) \nonumber \\
& \qquad \qquad \qquad \qquad \qquad \qquad \qquad \qquad \qquad \qquad \qquad +  g_{4}^{(3)}(\vec{k}_1,\vec{k}_2, \vec{k}_3, \vec{k}_4) \, S_4^{(3)}(\vec{k}_1, \vec{k}_2, \vec{k}_3, \vec{k}_4) \bigg)\,.
\end{align}
\end{widetext}
Let us specify the $g_{2,0}$, $g_{2,1}$, $g_4^{(1)}$, and $g_4^{(3)}$ kernels.  Defining
\begin{align}
\mathcal{G}_1(\vec{k}_1,\vec{k}_2,\vec{k}_3,\vec{k}_4) &:= 2\big(g_{2,0}(\vec{k}_1) - g_{2,0}(\vec{k}_2)
- g_{2,0}(\vec{k}_3) - g_{2,0}(\vec{k}_4)\big), \nonumber \\
\mathcal{G}_3(\vec{k}_1,\vec{k}_2,\vec{k}_3,\vec{k}_4) &:= 2\big(g_{2,0}(\vec{k}_1) + g_{2,0}(\vec{k}_2)
+ g_{2,0}(\vec{k}_3) - g_{2,0}(\vec{k}_4)\big)\, \nonumber.
\end{align}
and further defining
\begin{align}
\widetilde{g}_4^{(j)}(\vec{k}_1, \vec{k}_2, \vec{k}_3, \vec{k}_4) &:= \frac{e^{- \mathcal{G}_j(\vec{k}_1,\vec{k}_2,\vec{k}_3,\vec{k}_4)}-1}{\mathcal{G}_j(\vec{k}_1,\vec{k}_2,\vec{k}_3,\vec{k}_4)} \, g_4^{(j)}(\vec{k}_1, \vec{k}_2, \vec{k}_3, \vec{k}_4)
\end{align}
with $j=1,3$, then the functions in $U$ above are given by
\begin{align}
\label{sol00}
g_{2,0}(\vec{k}) = \frac{1}{4} \log \left(\frac{\Omega(\vec{k})}{\omega_k}\right)
\end{align}
\begin{align}
\label{sol01}
\begin{split}
g_{2,1}(\vec{k}_1) &= - \frac{1}{\omega_{k_1}^2} \Big(\frac{e^{-2u} (\delta m^2/\lambda)}{8}
\\&\;\;\;\;\;\;\;
+ \frac{e^{(d-3)u}}{32} \frac{1}{(2\pi)^{d}}\int d^d \vec{k}_2 \, \frac{1}{\omega_{k_1}\!+\!\omega_{k_2}} \Big)
\end{split}
\end{align}
\begin{align}
\begin{split}
\label{sol02}
&\widetilde{g}_{4}^{(1)}(\vec{k}_1,\vec{k}_2,\vec{k}_3,\vec{k}_4)
\\
&\;\;\;\; = \frac{1}{96}\frac{e^{(d-3)u}}{(2\pi)^{d}}\,\frac{1}{\omega_{k_2} \omega_{k_3} \omega_{k_4}(\omega_{k_1}\!+\!\omega_{k_2}\!+\!\omega_{k_3}\!+\!\omega_{k_4})}
\end{split}
\\
\begin{split}
\label{sol03}
&\widetilde{g}_{4}^{(3)}(\vec{k}_1,\vec{k}_2,\vec{k}_3,\vec{k}_4)
\\
&\;\;\;\; = \frac{1}{32}\frac{e^{(d-3)u}}{(2\pi)^{d}}\,\frac{1}{\omega_{k_4}(\omega_{k_1}\!+\!\omega_{k_2}\!+\!\omega_{k_3}\!+\!\omega_{k_4})}\,\,.
\end{split}
\end{align}
The unitary that we have constructed is not the unique unitary satisfying $U|\Psi_0\rangle = |\Psi(\Lambda e^u)\rangle + \mathcal{O}(\lambda^2)$.  For instance, if we have any unitaries $U_1, U_2$ satisfying
\begin{align}
U_1 |\Psi_0\rangle &= |\Psi_0\rangle \, + \mathcal{O}(\lambda^2) \\
U_2 |\Psi(\Lambda e^u)\rangle &= |\Psi(\Lambda e^u)\rangle \, + \mathcal{O}(\lambda^2) 
\end{align}
then we have
\begin{equation}
U_2 U U_1 |\Psi_0\rangle = |\Psi(\Lambda e^u)\rangle \, + \mathcal{O}(\lambda^2)\,.
\end{equation}
Therefore, $U_2  U U_1$ is also a viable unitary for our purposes.  It is in fact possible to construct the most general unitary mapping $|\Psi_0\rangle$ to $|\Psi(\Lambda e^u)\rangle$ up to $\mathcal{O}(\lambda^2)$ corrections, but we will not do so here.  It will suffice to consider our particular unitary $U$ in Eqn.~\eqref{unitary1}.

\section{cMERA for Weakly Interacting Fields}

cMERA is a variational ansatz for the ground states of field theories.  The ansatz, which lives in the UV, is constructed by building up entanglement hierarchically from an unentangled, scale-invariant IR state.  Concretely, consider the IR state $|\Omega\rangle$ which has the form
\begin{align}
\label{IRstate1}
\langle \phi | \Omega\rangle &= \mathcal{N} \, \exp\left(- \frac{1}{2} \int d^d \vec{x} \, \phi(\vec{x}) \, M \, \phi(\vec{x}) \right) \nonumber \\
&= \mathcal{N} \, \prod_{\vec{x}} \, \exp\left(- \frac{1}{2} \,d^d \vec{x} \, \phi(\vec{x}) \, M \, \phi(\vec{x}) \right)
\end{align}
for some constant $M$. Notice that $|\Omega\rangle$ is separable (i.e., spatially unentangled) and is scale-invariant with respect to spatial dilatations, i.e. $e^{-i u L}|\Omega\rangle = |\Omega\rangle$ with $L$ being the spatial dilatation operator
\begin{align}
L &= - \frac{1}{2} \int d^d \vec{x} \, \bigg(\widehat{\pi}(x) \left(\vec{x} \cdot \vec{\nabla}\widehat{\phi}(\vec{x})\right) + \left(\vec{x} \cdot \vec{\nabla}\widehat{\phi}(\vec{x})\right) \widehat{\pi}(\vec{x}) \nonumber \\
& \qquad \qquad \qquad \qquad \qquad \,\,+ \frac{d}{2} \, \widehat{\phi}(\vec{x}) \widehat{\pi}(\vec{x}) + \frac{d}{2} \, \widehat{\pi}(\vec{x}) \widehat{\phi}(\vec{x}) \bigg)\,.
\end{align}
The cMERA ansatz takes the form of the path-ordered exponential
\begin{equation}
\label{cMERAansatz1}
|\Psi_{\text{cMERA}}\rangle = \mathcal{P}_s \, \exp\left(-i \int_{u_{\text{IR}}}^{u_{\text{UV}}} ds \, (K(s) + L)\right)|\Omega\rangle
\end{equation}
where $K(s)$ is called the entangler, which contains free parameters that we variationally optimize by minimizing $\langle\Psi_{\text{cMERA}}|H_{\text{UV}}|\Psi_{\text{cMERA}}\rangle$ for some UV Hamiltonian.  For concreteness, we let $u_{\text{IR}} = -\infty$ and $u_{\text{UV}} = 0$.  Eqn.~\eqref{cMERAansatz1} has a straightforward interpretation: $K(s)$ creates correlations at a distance scale $\sim \,\Lambda^{-1} \exp(-s)$, for $-\infty < s \leq 0$.  Or equivalently, in momentum space, $K(s)$ creates correlations at a momentum scale $\sim \, \Lambda\,\exp(s)$ for $-\infty < s \leq 0$.

If we want to capture the correlations of $|\Psi_{\text{cMERA}}\rangle$ in Eqn.~\eqref{cMERAansatz1} renormalized down to the momentum scale $\Lambda e^u$ (i.e., distance scale $\Lambda^{-1} e^{-u}$) for $-\infty \leq u \leq 0$, then we would write
\begin{equation}
\label{cMERAansatz2}
|\Psi_{\text{cMERA}}(\Lambda e^u)\rangle = \mathcal{P}_s \, \exp\left(-i \int_{-\infty}^{u} ds \, (K(s) + L)\right) |\Omega\rangle\,.
\end{equation}

Even though we have cast cMERA as a variational ansatz, all previously known applications have been for the ground states of free bosonic or free fermionic theories \cite{cMERA1, cMERA2, cMERA3, cMERA4}.  The ground state of a free theory is a Gaussian wavefunctional, and one can find $K(s)$ exactly so that $|\Psi_{\text{cMERA}}(\Lambda)\rangle$ agrees with a free ground state in the UV.  Even though mean field theory has been applied to cMERA \cite{CMM1}, this approach has limited utility.

One complication with computing $K(s)$ for interacting theories is that their RG flows are non-trivial, unless the theory is a CFT.  Necessarily, $K(s)$ must encode information about the RG flow, and so will have a more complicated form vis-\`{a}-vis free theories.

In this section, we will use quantum circuit perturbation theory to construct a \textit{local} $K(s)$ such that the corresponding cMERA state agrees with the 1-loop UV ground state of $\varphi^4$ theory.  Additionally, our cMERA state will have $|\Psi_{\text{cMERA}}(\Lambda e^u)\rangle$ equal with the 1-loop Wilsonian renormalized ground state of $\varphi^4$ theory at \textit{all} intermediate RG scales $\Lambda e^u$.  This establishes a direct correspondence between cMERA circuits with local entanglers, and Wilsonian RG on spatial momentum modes.

Before proceeding to $\varphi^4$ theory, we will first compute $K(s)$ for the ground state of a free massive scalar field theory, such that $|\Psi_{\text{cMERA}}(\Lambda e^u)\rangle$ equals the Wilsonian renormalized ground state at all intermediate RG scales.  This is distinct from previous work, which only required that the cMERA state agree with a desired UV ground state \cite{cMERA1, cMERA2, cMERA3, cMERA4}.

The \textit{exact} Wilsonian renormalized Hamiltonian for a massive scalar field theory is given by Eqn.'s~\eqref{phi4Ham1loop} and~\eqref{deltaM1} with $\lambda = 0$. The ground state renormalized to scale $\Lambda e^u$ is given by
\begin{equation}
\langle \phi | \Psi_0(\Lambda e^u)\rangle = \mathcal{N} \, e^{- \frac{1}{2} \int d^d \vec{k}\,\theta(1-|\vec{k}|/\Lambda)\,  \phi_{\vec{k}} \, \sqrt{\vec{k}^2 + e^{-2u} m^2} \, \phi_{-\vec{k}} }
\end{equation}
where $\theta(z)$ is an analytic (and thus smooth) version of the Heaviside step function, for instance a sigmoid.  Thus, $\theta(1-|\vec{k}|/\Lambda)$ provides a smooth cutoff at $|\vec{k}| = \Lambda$.  (Recall that when we perform Wilsonian RG down to scale $\Lambda e^u$, we rescale the momenta so that the cutoff is set back to $\Lambda$, and also renormalize the fields to put the kinetic term of the Hamiltonian in a canonical form.)  Letting $M = \sqrt{\Lambda^2 + m^2}$ in Eqn.~\eqref{IRstate1}, we find that the desired entangler is
\begin{widetext}
\vskip-.5cm
\begin{equation}
\label{entanglerMassive1}
K(s) = \int d^d \vec{k} \,  \left[\frac{1}{4} \, \theta(1-|\vec{k}|/\Lambda) - \frac{1}{8} \log\left(\frac{\vec{k}^2 + e^{-2s}m^2}{\Lambda^2 + m^2}\right) \frac{|\vec{k}|}{\Lambda} \, \theta'(1-|\vec{k}|/\Lambda)\right] \, \left( \phi_{\vec{k}}\,\pi_{-\vec{k}} + \pi_{\vec{k}}\, \phi_{-\vec{k}}\right)
\end{equation}
\end{widetext}
Our answer has several interesting features in position space.  The Fourier transform of $\frac{1}{4} \, \theta(1-|\vec{k}|/\Lambda)$ is a function localized at the origin with width $1/\Lambda$ which leads to correlations at scale $\Lambda e^s$ in the cMERA state.  The Fourier transform of $-\frac{1}{8} \log\left(\frac{\vec{k}^2 + e^{-2s}m^2}{\Lambda^2 + m^2}\right) \frac{|\vec{k}|}{\Lambda} \, \theta'(1-|\vec{k}|/\Lambda)$ is localized at the origin with width $1/(e^{-s} m)$ (i.e., the inverse renormalized mass scale), but it has essentially zero amplitude unless $1/m \lesssim \Lambda^{-1}\,e^{-s}$.  This result means that we can only see the effect of the mass $m$ of the UV theory if we probe distance scales around $\sim 1/m$ or larger.  Probing shorter distance scales essentially only touches massless modes.

Now we construct the entangler $K(s)$ for the ground state of $\varphi^4$ theory, such that $|\Psi_{\text{cMERA}}(\Lambda e^u)\rangle$ equals the Wilsonian renormalized ground state at all intermediate RG scales to 1-loop.  Recall that the desired ground state $|\Psi(\Lambda e^u)\rangle$ is given by Eqn.~\eqref{phi4gs1} above.  Letting
\begin{align}
\begin{split}
M &= \sqrt{\Lambda^2 + m^2} + \lambda \bigg(\frac{1}{2}\,(\delta m^2/\lambda) \, \frac{1}{M}
\\&\qquad
-\frac{4}{(2\pi)^d}\int^\Lambda d^d \vec{q} \, \frac{1}{M(M+\sqrt{q^2+m^2})} 
 \bigg)
\end{split}
\end{align}
\begin{figure*}[ht]
\centering
\includegraphics[width=7.4cm]{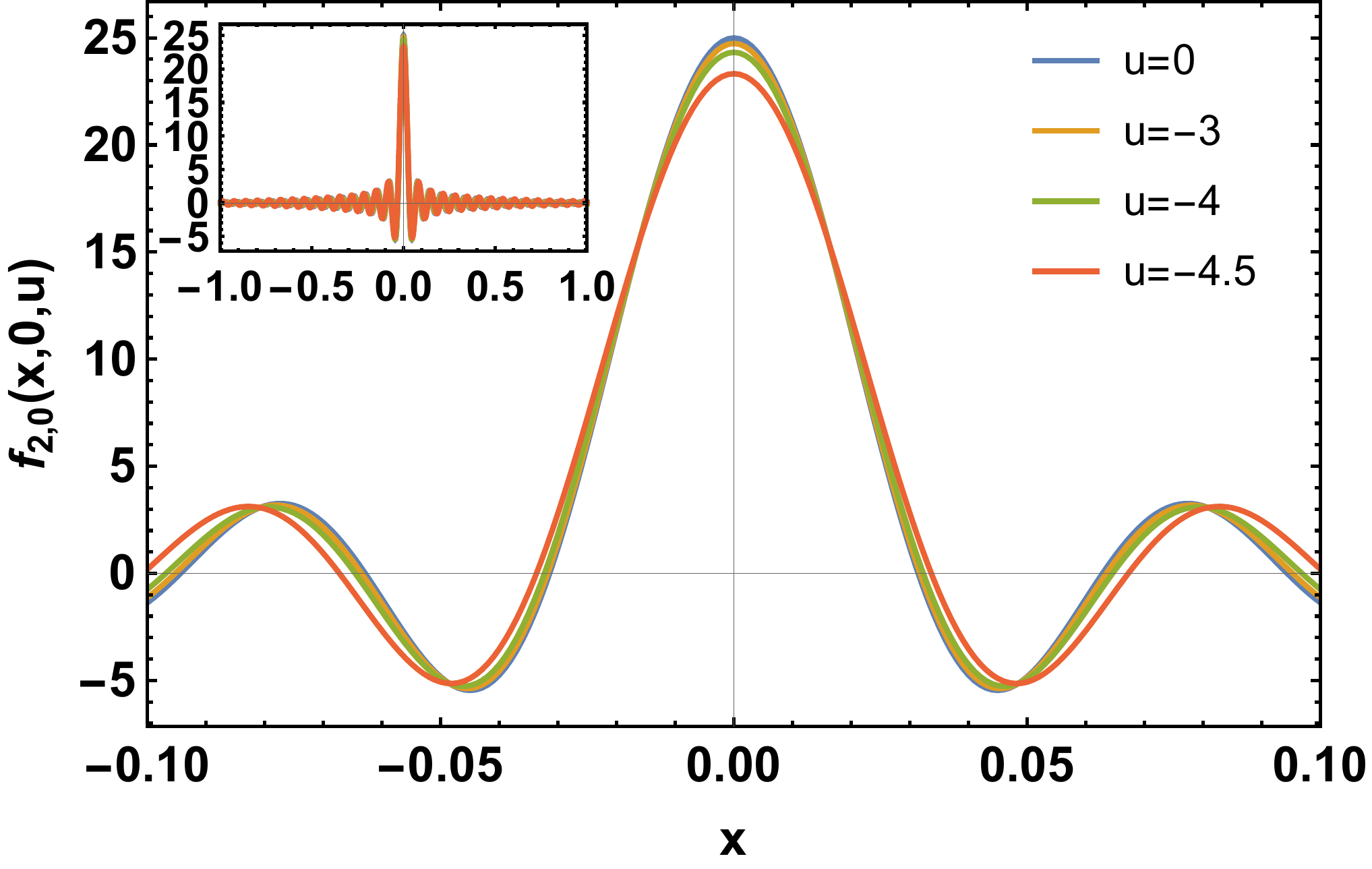}
\hspace{0mm}
\includegraphics[width=7.4cm]{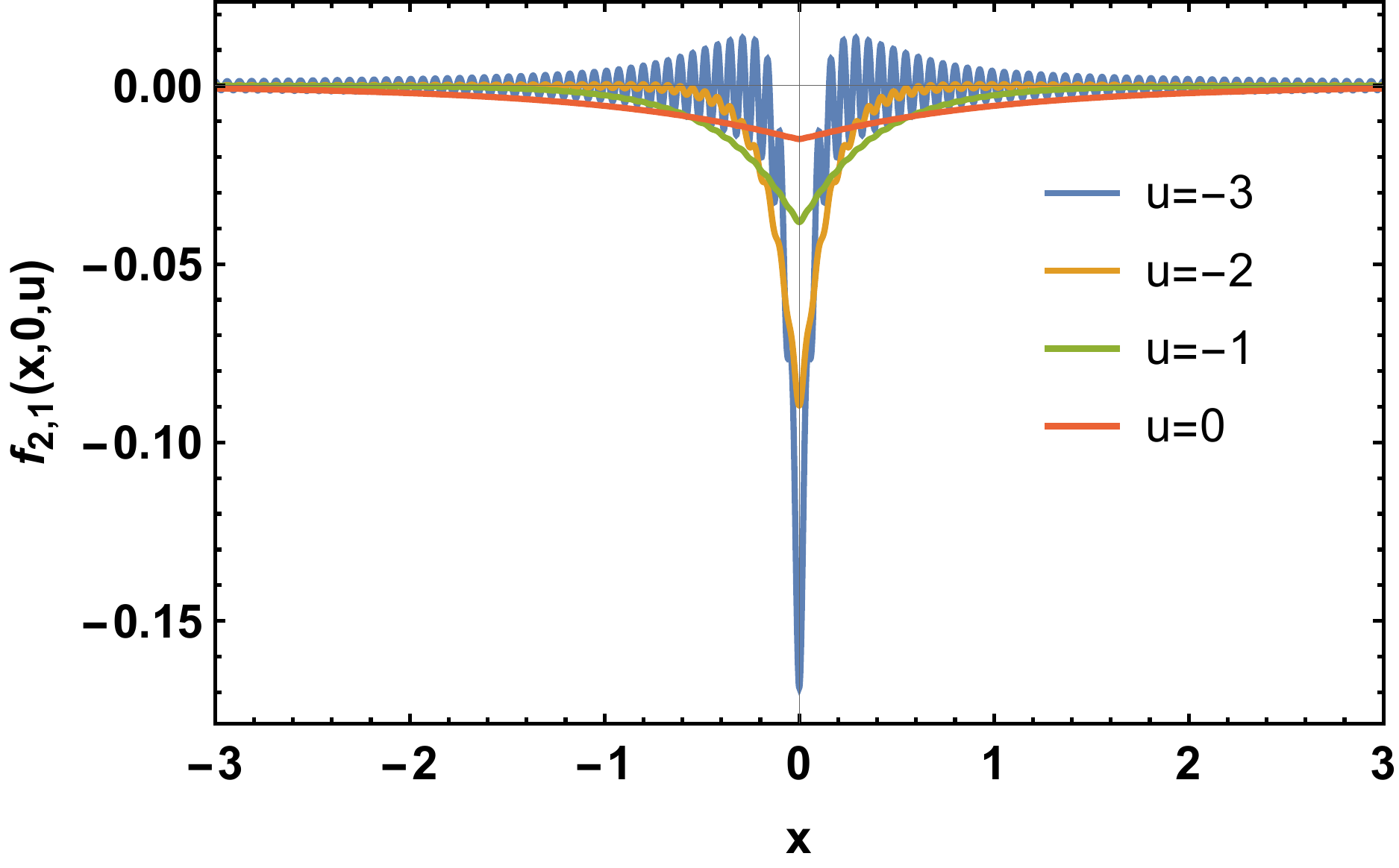}
\\
\hspace{1mm}
\includegraphics[width=7.3cm]{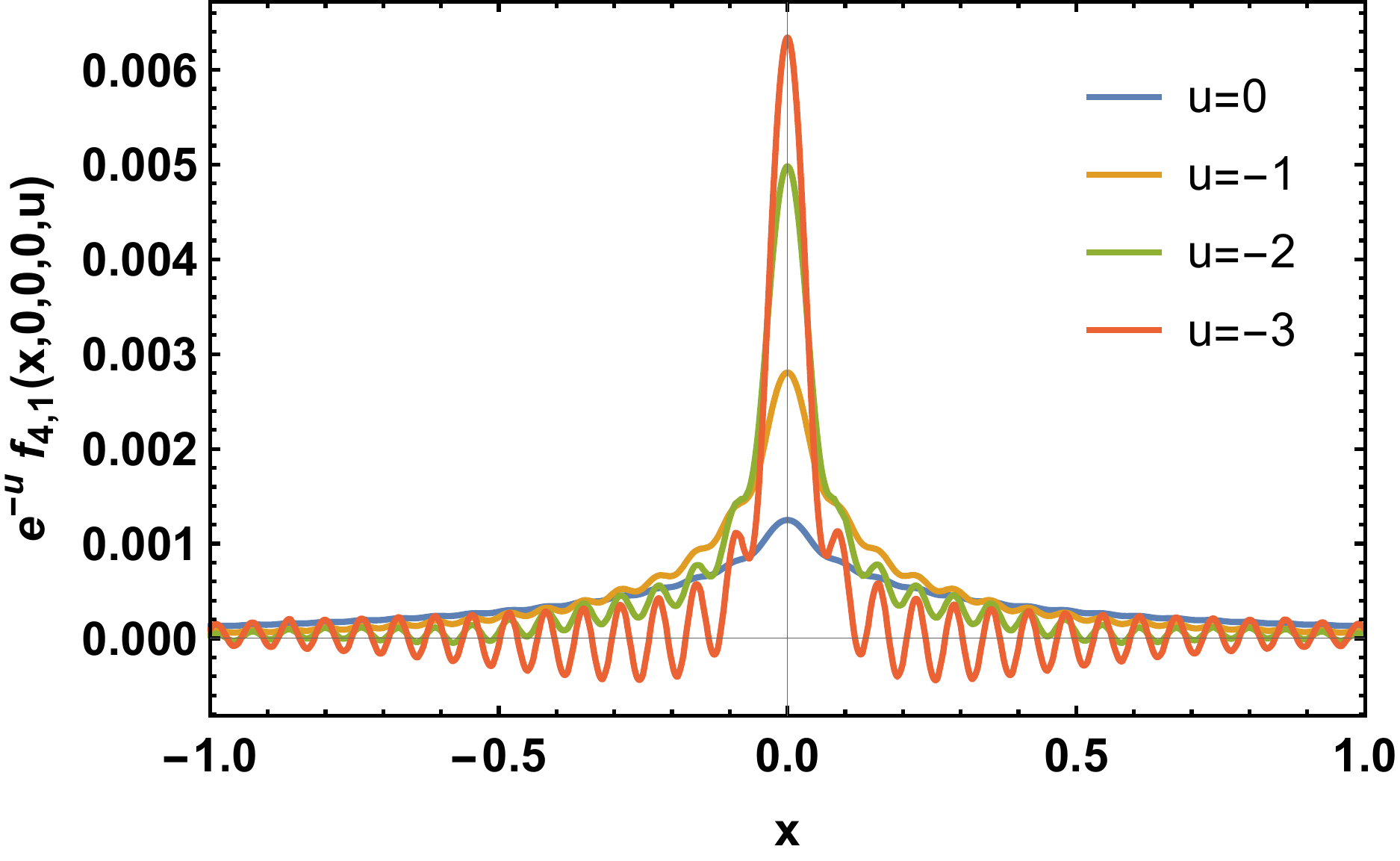}
\hspace{1mm}
\includegraphics[width=7.3cm]{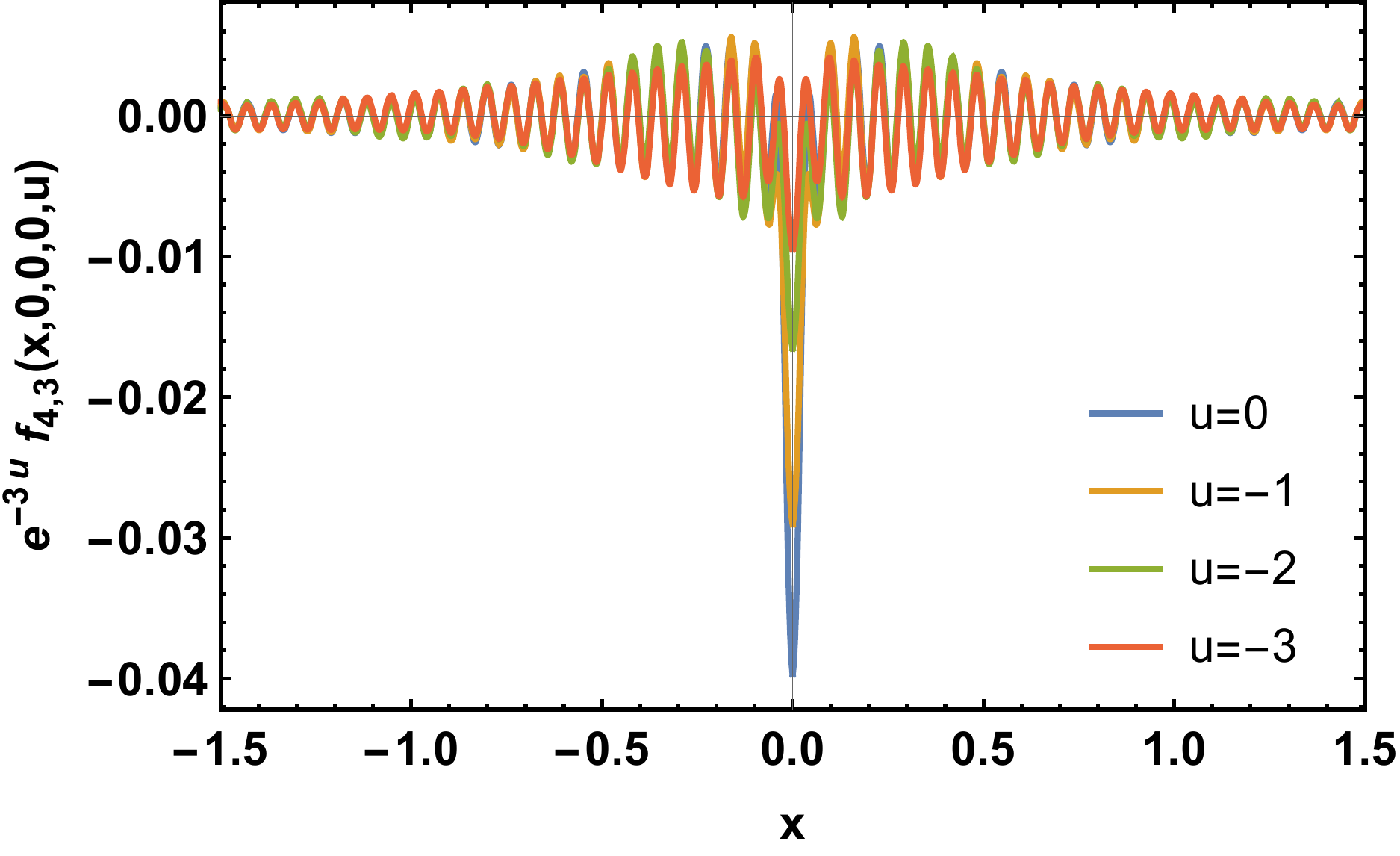}
\caption{The kernels comprising the entangler in position space at different values of $u$. Here we have set $m=1$ and $\Lambda=100$.}
\label{fig:kernels}
\end{figure*}

in Eqn.~\eqref{IRstate1}, we find that the 1-loop entangler is
\begin{widetext}
\vskip-.5cm
\begin{align}
\label{untildeKform}
&K(s) = \int d^d \vec{k}_1 \, d^d \vec{k}_2 \, \delta^{(d)}(\vec{k}_1 + \vec{k}_2)\,f_{2,0}(\vec{k}_1 \, ; \, s) \, S_2^{(1)}(\vec{k}_1, \vec{k}_2) + \lambda \, \int d^d \vec{k}_1 \, d^d \vec{k}_2 \, \delta^{(d)}(\vec{k}_1 + \vec{k}_2)\,f_{2,1}(\vec{k}_1 \, ; \, s) \, S_2^{(1)}(\vec{k}_1, \vec{k}_2) \nonumber \\
& + \lambda \, e^{ds}\!\int d^d \textbf{k} \, \delta^{(d)}(\vec{k}_1\!+\!\vec{k}_2\!+\!\vec{k}_3\!+\!\vec{k}_4) \bigg( f_{4}^{(1)}(\vec{k}_1,\vec{k}_2,\vec{k}_3,\vec{k}_4 \, ; \, s) \, S_4^{(1)}(\vec{k}_1, \vec{k}_2, \vec{k}_3, \vec{k}_4) +  f_{4}^{(3)}(\vec{k}_1, \vec{k}_2, \vec{k}_3, \vec{k}_4 \, ; \, s) \, S_4^{(3)}(\vec{k}_1, \vec{k}_2, \vec{k}_3, \vec{k}_4) \bigg)\,,
\end{align}
\end{widetext}
where $f_{2,0}$, $f_{2,1}$ are given by
\begin{align}
\label{eq:f2i}
f_{2,i}(e^{-u}\vec{k}\,;\,u) &= \frac{d}{du} \left[\theta(1-|\vec{k}|/\Lambda e^u)  \, g_{2,i}(e^{-u}\vec{k} \, ; \, u)\right]
\end{align}
for $i=0,1$, and $f_{4}^{(1)}$, $f_{4}^{(3)}$ are given by
\begin{widetext}
\vskip-.5cm
\begin{align}
\label{f41sol2}
&f_4^{(j)}(\vec{k}_1, \vec{k}_2, \vec{k}_3, \vec{k}_4 \, ; \, u) = e^{-(d+1)u}\Bigg\{e^u\left(d - \frac{\partial \mathcal{F}_j(s,u)}{\partial u}\right) \, \widetilde{g}_4^{(j)}(\vec{k}_1, \vec{k}_2, \vec{k}_3, \vec{k}_4 \, ; \, u) - e^u \sum_{j=1}^4 \vec{k}_j \cdot \frac{\partial}{\partial \vec{k}_j}\, \widetilde{g}_4^{(j)}(\vec{k}_1, \vec{k}_2, \vec{k}_3, \vec{k}_4 \, ; \, u) \nonumber \\
& \qquad \qquad \qquad + e^u \frac{\partial}{\partial u} \widetilde{g}_4^{(j)}(\vec{k}_1, \vec{k}_2, \vec{k}_3, \vec{k}_4 \, ; \, u) - \widetilde{g}_{4}^{(j)} (\vec{k}_1, \vec{k}_2, \vec{k}_3, \vec{k}_4 \, ; \, u) \, \sum_{\ell=1}^4 \frac{|\vec{k}_\ell|}{\Lambda} \frac{\theta'(1-|\vec{k}_p|/\Lambda)}{\theta(1-|\vec{k}_\ell|/\Lambda)}\Bigg\} 
 \prod_{p=1}^4 \theta(1-|\vec{k}_p|/\Lambda)
\end{align}
\end{widetext}
with $j=1,3$, and $\mathcal{F}_1(s,u)$ and $\mathcal{F}_3(s,u)$ defined by
\begin{align*}
\mathcal{F}_1(s,u) &:= 2 \int_s^u dt \, \big(f_{2,0}(e^{-t}\vec{k}_1 \, ; \, t) - f_{2,0}(e^{-t}\vec{k}_2\, ; \, t) \nonumber \\
& \qquad \qquad \quad - f_{2,0}(e^{-t}\vec{k}_3\, ; \, t) - f_{2,0}(e^{-t}\vec{k}_4\, ; \, t) \big) \\
\mathcal{F}_3(s,u) &:= 2 \int_s^u dt \, \big(f_{2,0}(e^{-t}\vec{k}_1\, ; \, t) + f_{2,0}(e^{-t}\vec{k}_2\, ; \, t) \nonumber \\
& \qquad \qquad \quad + f_{2,0}(e^{-t}\vec{k}_3\, ; \, t) - f_{2,0}(e^{-t}\vec{k}_4\, ; \, t) \big) \,.
\end{align*}

The result is unwieldy, but has several remarkable properties.  Most importantly, some Fourier analysis shows that all of the kernels in $K(s)$ decay at worst exponentially in position space, with decay constant $\sim 1/(e^{-s} m)$ (i.e., the inverse renormalized mass scale).  More specifically, each kernel decays at worst exponentially in the the distance $|\vec{x}_i - \vec{x}_j|$ between any two spatial position vectors $\vec{x}_i$, $\vec{x}_j$ appearing in the kernel.  The locality of the kernels arises due to the poles of the massive propagator, as well as the relativistic dispersion relation for massive fields $\omega_{k} = \sqrt{\vec{k}^2 + e^{-2s} m^2}$.  (Here we have renormalized the mass to momentum scale $\Lambda e^{s}$.)  The mass terms shift the poles of the kernels off the real axis (or axes, for the multidimensional integrals) so we can analytically deform the contour of the Fourier transform to achieve exponential decay in position space.

We have plotted spatial slices of the kernels in Figure \ref{fig:kernels}.  The implies that the entangler is exponentially local in position space, and thus for the case at hand, Wilsonian RG on spatial momentum modes can be re-expressed in terms of a local cMERA circuit to 1-loop in perturbation theory.  This creates a direct link between more standard momentum space Wilsonian RG, and cMERA tensor networks.  Since the exponential position space locality of the kernels only depends on the pole structure of the propagator and the relativistic dispersion relation for massive fields, we expect analogous results to hold to higher loops in $\varphi^4$ theory, and for other massive theories.  For massless theories, the kernels will have a weaker decay due to the altered pole structure of the propagator and the relativistic dispersion relation for massless fields.

Our result for $K(s)$ for $\varphi^4$ theory has the property that the mass-dependent parts of the entangler only ``activate'' at distance scales greater than $\sim 1/m$, which is analogous to the free theory result above.  

\section{Lessons for Numerics}

The ultimate goal of cMERA is to provide a robust numerical ansatz for the ground state of an interacting quantum field theory.  We have focused on developing machinery for perturbative calculations to bring cMERA into the new territory of interacting field theories, albeit weakly interacting.  However, we can use insights from our calculations to construct an ansatz which may be viable for numerical variational calculations.

We construct an entangler
\begin{align}
\hspace{-3mm}
\begin{split}
& K(s) = \sum_{j_1} \int_{1/\Lambda} d^d \vec{x}_1 \, f_{j_1}(\vec{x}_1 \, ; \, u) \, \mathcal{O}_{j_1}(\vec{x}_1)
\\
&\;\;+ \sum_{j_2} \int_{1/\Lambda} d^d \vec{x}_1 \, d^d \vec{x}_2 \, f_{j_2}(\vec{x}_1, \vec{x}_2\, ; \, u) \, \mathcal{O}_{j_2}(\vec{x}_1, \vec{x}_2) 
+ \cdots
\\
&\;\;
+ \sum_{j_n} \int_{1/\Lambda} d^d \vec{x}_1 \cdots d^d \vec{x}_n \, f_{j_n}(\vec{x}_1,...,\vec{x}_n\, ; \, u) \, \mathcal{O}_{j_n}(\vec{x}_1,...,\vec{x}_n)\,
\end{split}
\end{align}
where the position space integrals are cutoff from below at scale $1/\Lambda$.  We take $f_{i_1},...,f_{i_n}$ to be functions of a specified form, but with undetermined parameters that we can tune and optimize.  We approximate the cMERA circuit $\mathcal{P}_s \exp\left(- i \int_{u_{\text{IR}}}^0 ds \, (K(s) + L) \right)$ by
\begin{align}
\hspace{-3mm}
\left[e^{- i \,\Delta u \,(K(u_{\text{IR}}) + L)} \,\, e^{- i \,\Delta u \,(K(u_{\text{IR}} + \Delta u) + L)}  \, \cdots \, e^{- i \,\Delta u \,(K(0) + L)} \right]_{T}
\end{align}
where $\Delta u := -u_{\text{IR}}/N$ for some positive integer $N$, and $[ \cdots ]_{T}$ denotes that we truncate the terms inside the bracket at order $\mathcal{O}((\Delta u)^T)$.  Our cMERA ansatz is
\begin{align}
\begin{split}
|\Psi_{\text{cMERA}}\rangle &:= \Big[e^{- i \,\Delta u \,(K(u_{\text{IR}}) + L)} \times 
\\  & 
 e^{- i \,\Delta u \,(K(u_{\text{IR}} + \Delta u) + L)}\cdots \, e^{- i \,\Delta u \,(K(0) + L)} \Big]_{T}\, |\Omega\rangle
\end{split}
\end{align}
which depends on the functions $f_{i_1},...,f_{i_n}$.  To utilize this ansatz, we consider a UV Hamiltonian $H_{\text{UV}}$ and perform the numerical minimization
\begin{equation}
\min_{f_{i_1},...,f_{i_n}} \frac{\langle \Psi_{\text{cMERA}}|H_{\text{UV}}|\Psi_{\text{cMERA}}\rangle}{\langle \Psi_{\text{cMERA}} | \Psi_{\text{cMERA}} \rangle}
\end{equation}
where the denominator is required since $|\Psi_{\text{cMERA}}\rangle$ is not normalized as given.  Our calculations suggest that a good way of parametrizing the $f_{i_1},...,f_{i_n}$ is in terms of Sine-Gaussian wavelets which only depend on the differences of coordinates $|\vec{x}_i - \vec{x}_j|$.  For instance, we might parametrize a kernel $f(\vec{x}_1,\vec{x}_2)$ by
\begin{align}
\label{paramkern1}
&f(\vec{x}_1,\vec{x}_2 \, ; \, \{a_j, b_j, c_j, d_j, \phi_j\}) = \\
& \qquad \quad \sum_{j} a_j \, e^{- b_j^2 \, |\vec{x}_1 - \vec{x}_2|^2 + c_j \, |\vec{x}_1 - \vec{x}_2|} \cos(d_j \, |\vec{x}_1 - \vec{x}_2| + \phi_j) \nonumber
\end{align}
which is the form of the sum of the real parts of Gabor wavelets.  A kernel $f(\vec{x}_1,\vec{x}_2, \vec{x}_3, \vec{x}_4)$ might be parametrized similarly by
\begin{align}
\label{paramkern2}
& f(\vec{x}_1,\vec{x}_2, \vec{x}_3, \vec{x}_4 \, ; \, \{a_j, \textbf{B}_j, \textbf{c}_j, \textbf{d}_j, \boldsymbol{\phi}_j\}) \\ &\qquad \qquad \qquad = \sum_{j} a_j \, e^{- \textbf{x}^T \textbf{B}_j \textbf{x} + \textbf{c}_j \cdot \textbf{x}} \cos(\textbf{d}_j \cdot \textbf{x} + \boldsymbol{\phi}_j) \nonumber 
\end{align}
where $\textbf{x} := ( \, |\vec{x}_1 - \vec{x}_2|\,,\,|\vec{x}_1 - \vec{x}_3|\,,\, |\vec{x}_1 - \vec{x}_4|\,,\,|\vec{x}_2 - \vec{x}_3|\,,\,|\vec{x}_2 - \vec{x}_4|\,,\,|\vec{x}_3 - \vec{x}_4|\,)$.  Here $\textbf{B}_j$ is a $6 \times 6$ matrix of parameters, and $\textbf{c}_j, \textbf{d}_j, \boldsymbol{\phi}_j$ are all $6$--dimensional vectors of parameters.  For non-CFT’s, the parameters in Eqn.'s~\eqref{paramkern1} and~\eqref{paramkern2} can depend on $u$, and thus have non-trivial dependence on the distance scale.

We envision that by parametrizing the kernels $f_{i_1},...,f_{i_n}$ in terms of appropriate Sine-Gaussian wavelets, it should be possible for cMERA to become a useful variational method for the ground states of CFT's as well as regular QFT's (for which there are additional parametric dependencies in the kernels).  In particular, the integrals and gradient descent procedure required to minimize
$$\langle \Psi_{\text{cMERA}}|H_{\text{UV}}|\Psi_{\text{cMERA}}\rangle/\langle \Psi_{\text{cMERA}} | \Psi_{\text{cMERA}} \rangle$$
over the parameters of Sine-Gaussian wavelets (or similar such wavelets) can be performed efficiently.
\vskip-.5cm
\section{Discussion}

We have shown that we can perturbatively construct a cMERA with a local entangler for the ground state of weakly interacting $\varphi^4$ theory.  Furthermore, the cMERA can be constructed to agree with Wilsonian RG on spatial momentum modes.  Our procedure is systematic, and should provide similar constructions for other QFT's.  In particular, we expect that cMERA kernels for other massive theories, given by our procedure, will also be exponentially localized in position space.  Furthermore, we expect that cMERA kernels for massless theories will also be localized in position space, but not exponentially.  We have also used our calculations to motivate a numerical approach to cMERA, which does not require field to be weakly interacting.

There are several interesting future directions.  First, it would be interesting to perform higher loop calculations, and to generalize the results to fermionic theories \cite{fermion} and gauge fields.  One can also compute the cMERA circuit for weakly interacting CFT's like the Wilson-Fisher fixed point.  It may also be possible to generalize our perturbative techniques to many-body spin systems, along the lines of \cite{Bridging1}.

Since tensor networks are intrinsically tied with entanglement properties of the quantum states they generate, a detailed study of the entanglement properties of weakly interacting cMERA circuits may yield new insights.  It would also be interesting to understand the connection to ``flow equations'' \cite{Wegner1, Stefan1, Stefan2} and various generalizations of holography \cite{holo1,holo2,holo3,holo4}.  One could also explore complexity for weakly interacting field theories, along the lines of \cite{complexity1, complexity2}. 
\vskip.2cm
\textit{Acknowledgements:} We would like to thank Chris Akers, Ignacio Cirac, William Donnelly, Patrick Hayden, Michal Heller, Javier Molina-Vilaplana, Mark Mueller, Tobias Osborne, Daniel Ranard, Tadashi Takayanagi, Frank Verstraete, and Guifr\'{e} Vidal for valuable conversations and feedback. We thank Felipe Hern\'{a}ndez for many discussions about the locality of the cMERA kernels in position space.  JC is supported by the Fannie and John Hertz Foundation and the Stanford Graduate Fellowship program. AM acknowledges support by the Alexander von Humboldt Foundation via a postdoctoral fellowship. AM also would like to thank ICTP for hospitality during the last stages of this work. AN has greatly profited from discussions with Farhad Ardalan on Wilsonian RG and also Vahid Karimipour and Niloofar Vardian on different aspects of tensor networks. AN also would like to thank CERN TH-Division and ICTP for hospitality during some stages of this work.

\nopagebreak
\end{document}